\documentclass[journal]{vgtc}           

\ifpdf
\pdfoutput=1\relax                   
\usepackage{graphicx}                
\DeclareGraphicsExtensions{.pdf,.png,.jpg,.jpeg} 
\else
\usepackage{graphicx}                
\DeclareGraphicsExtensions{.eps}     
\fi%

\graphicspath{{figures/}{pictures/}{images/}{./}} 

\usepackage{microtype}                 
\PassOptionsToPackage{warn}{textcomp}  
\usepackage{textcomp}                  
\usepackage{mathptmx}                  
\usepackage{times}                     
\usepackage{cite}                      
\usepackage{tabu}                      
\usepackage{booktabs}                  
\usepackage{microtype}
\usepackage[none]{hyphenat}
\usepackage{amsmath}
\usepackage{amssymb}
\usepackage{xspace}
\usepackage{url}
\usepackage[mathscr]{euscript}
\usepackage{paralist}
\usepackage{color}
\usepackage{xcolor}
\usepackage{colortbl}
\usepackage{CJK}
\usepackage{soul}
\usepackage[utf8]{inputenc}
\usepackage{hyperref}
\definecolor{wxm}{rgb}{1, 0.58, 0}

\onlineid{1338}

\vgtccategory{Research}
\vgtcpapertype{Applications}

\newcommand{\techname}{\textit{HetVis}\xspace}
\title{HetVis: A Visual Analysis Approach for Identifying\\Data Heterogeneity in Horizontal Federated Learning}

\author{Xumeng Wang, Wei Chen, Jiazhi Xia, Zhen Wen, Rongchen Zhu, Tobias Schreck}
\authorfooter{
\item
Xumeng Wang is with TMCC, CS, Nankai University. E-mail: wangxumeng@nankai.edu.cn. 
\item
Wei Chen, Zhen Wen, and Rongchen Zhu are with State Key Lab of CAD\&CG, Zhejiang University. E-mail: \{chenvis, whenzhen, zrcccrz\}@zju.edu.cn. 
\item
Wei Chen is also with Laboratory of Art and Archaeology Image (Zhejiang University), Ministry of Education, China. 
\item
Jiazhi Xia is with School of Computer Science and Engineering, Central South University. E-mail: xiajiazhi@csu.edu.cn.
\item
Tobias Schreck is with Graz University of Technology, Austria. E-mail: tobias.schreck@cgv.tugraz.at.
\item
Wei Chen and Jiazhi Xia are the corresponding authors.
}

\abstract{
Horizontal federated learning (HFL) enables distributed clients to train a shared model and keep their data privacy. In training high-quality HFL models, the data heterogeneity among clients is one of the major concerns. However, due to the security issue and the complexity of deep learning models, it is challenging to investigate data heterogeneity across different clients. To address this issue, based on a requirement analysis we developed a visual analytics tool, HetVis, for participating clients to explore data heterogeneity. We identify data heterogeneity through comparing prediction behaviors of the global federated model and the stand-alone model trained with local data. Then, a context-aware clustering of the inconsistent records is done, to provide a summary of data heterogeneity. Combining with the proposed comparison techniques, we develop a novel set of visualizations to identify heterogeneity issues in HFL. We designed three case studies to introduce how HetVis can assist client analysts in understanding different types of heterogeneity issues. Expert reviews and a comparative study demonstrate the effectiveness of HetVis.

} 


\CCScatlist{ 
\CCScat{K.6.1}{Management of Computing and Information Systems}%
{Project and People Management}{Life Cycle};
\CCScat{K.7.m}{The Computing Profession}{Miscellaneous}{Ethics}
}

\teaser{
   \centering
   \includegraphics[width=\columnwidth]{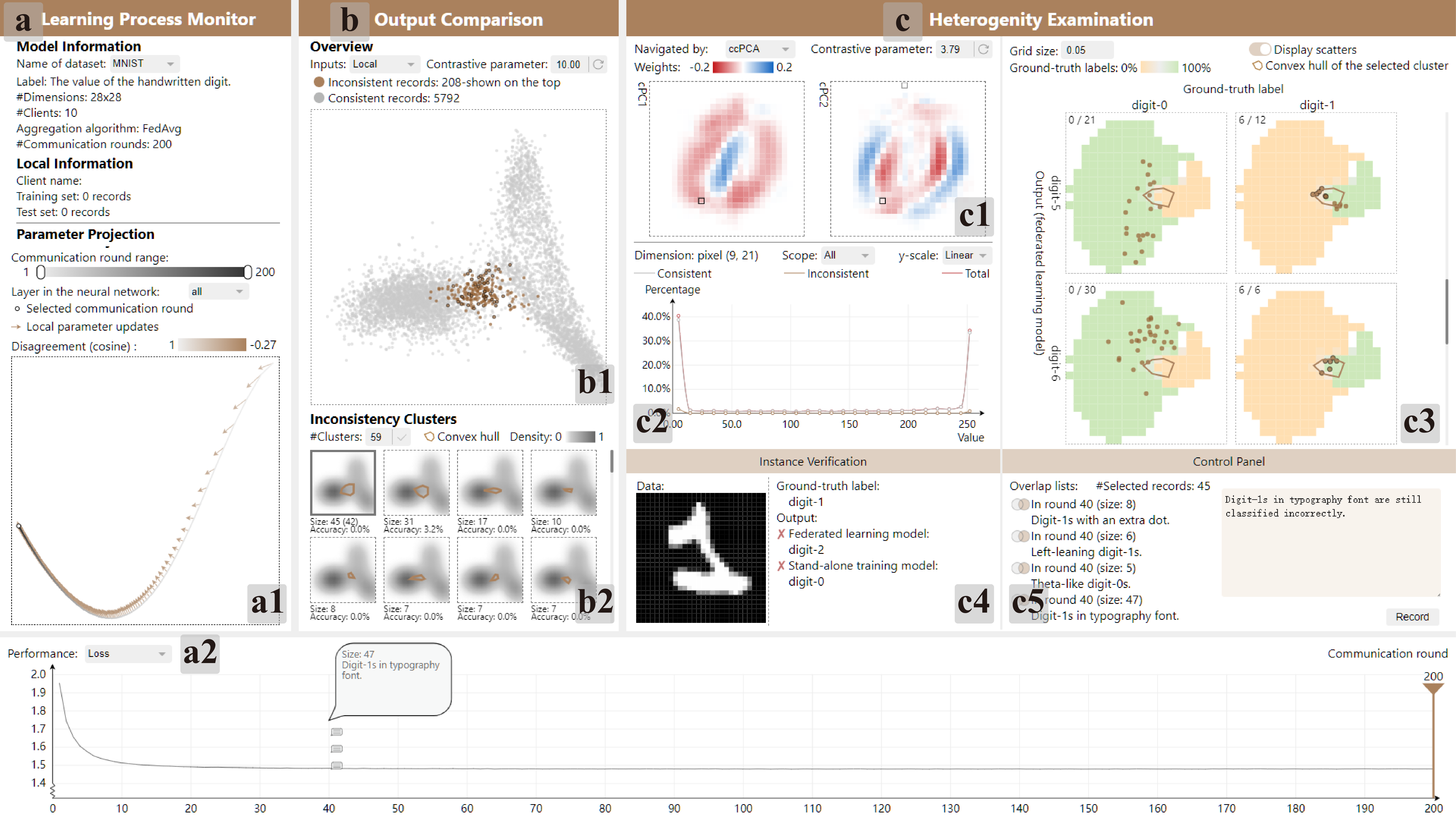}
   \caption{The interface of \techname for analyzing the heterogeneity issues among Federated Learning cooperation from the perspective of a participating client. (a) The module of federated learning process observation consists of an information panel introducing the federated model and the local client, the parameter projection view depicting the evolution of the disagreements between the local client and the integrated model, and a performance monitor view recording model performance and users' annotations. (b) The module of model output comparison identifies and clusters inconsistent records between the output of the federated learning model and the stand-alone training model. (c) The module of heterogeneity examination allows users to analyze an inconsistent cluster by observing the distribution and inspecting instances. Findings can be annotated to the cluster in the control panel.}
   \label{fig:tea}
}

\vgtcinsertpkg

\begin{document}
\firstsection{Introduction}

\maketitle



In the big data era, data is often distributed among many isolated data owners. It is an urgent and challenging need for the data owners to benefit from data integration, e.g., to train high-quality models, while also respecting privacy concerns. To satisfy this requirement, federated learning~\cite{yang2019federated} (FL) keeps data locally in clients when training a shared model. Only encrypted parameters are shared with other clients. Especially, horizontal federated learning (HFL), which integrates data from the same feature space but  distributed at different clients, has been widely used in privacy-aware applications, like healthcare and mobile service~\cite{li2020review, xu2020federated}.

A key and common challenge of HFL is the \emph{heterogeneity} of data distribution. The efficiency of current HFL techniques depends on the assumption that data distribution in different clients is independent and identically distributed (IID). However, the IID assumption usually does not hold. Non-IID data is widely used in model training to increase sample size and leads to difficulties in convergence~\cite{li2020federated, yang2020heterogeneity}. Figuring out their existence is necessary for applications of HFL. Specifically, data heterogeneity can exert both negative impacts and positive impacts on the accuracy of trained models. \textit{Statistical heterogeneity}, meaning that distributed data cover the same classes but different proportions, can be positive. For example, one client has insufficient samples in one class while the other clients have adequate samples in the same class. In this case, multiple clients are facilitated in the collaboration. \textit{Label heterogeneity} means that similar instances are labeled differently in different clients. The divergence in label settings usually affects model training negatively. Therefore, it is critical for participating clients to diagnose the data heterogeneity in the training process and fine-tune the model and data~\cite{yuan2021survey}. We identify two analysis goals in heterogeneity: (1) detecting the existence of data heterogeneity and (2) knowing the impacts of data heterogeneity. For example, hospitals collect patient data for disease analysis, and search engines record user logs for advertising recommendations.


However, analyzing the data heterogeneity in HFL is highly challenging. Traditional approaches identify heterogeneity through comparing data~\cite{rosenthal2002meta}. Due to the privacy issue, a client is prevented from accessing data owned by other clients directly. Direct comparison between local data and global data is infeasible.
The HFL model is the only intelligence fed back to the client by the cooperation. 
As it is trained with the global model with a HFL framework, the global data is learned by the HFL model.
There are two challenges in analyzing data heterogeneity based on the HFL model.
First, although a lot of works have been dedicated to understanding deep learning models by input and output data, the inverse workflow of understanding the unavailable training data by the model is still an open problem.
Second, the behavior of the HFL model depends on multiple factors, including the data and model architectures. How to distinguish the effect of training data from other factors is also challenging.
An encompassing  analysis workflow is still missing for HFL analysis.



To fill the above research gap, we present a visual analytics tool for client analysts to detect and understand data heterogeneity in HFL under the privacy limitation. We leverage a contrastive analysis approach to locate heterogeneous records and examine them. To overcome the privacy limitation, we propose to train a stand-alone model based on local data and compare it with the HFL model. Because data is learned by a model in the training process, the comparison between the two models discloses the heterogeneity between the local data and global data. We further propose a cluster analysis method based on rank-based distance measurement to distinguish the impacts from different group of heterogeneous records. 
We developed an interactive interface, called \techname, in conjunction with our analysis approach to monitor, explore, and understand the models and instances. A comparative study is implemented to show that the proposed cluster analysis method is more helpful for analyzing heterogeneity than hierarchical clustering. Three case studies were designed to demonstrate the effectiveness of our interface when analyzing different kinds of data heterogeneity. We also collected informative reviews from three experts in FL, confirming the effectiveness of \techname.

In summary, the contributions of this work are:
\begin{itemize}
    \item A novel visual analysis workflow that assists to analyze the heterogeneity in HFL models.
    \item  A context-aware clustering approach that hierarchically generates clusters of data which are inconsistent with others. 
    \item A visual system that integrates a suite of novel designs.
\end{itemize}

\section{Related Work}
In this section, related works are summarized from three aspects: heterogeneity in distributed learning, visualization for model diagnosis, and visualization for model comparison.

\subsection{Heterogeneity in Distributed Learning}

The problem of data heterogeneity has put forward challenges for integrating distributed data into joint or public knowledge. Related scenarios mainly include distributed learning and statistical analysis. 

In the first scenario, heterogeneity issues, including state heterogeneity~\cite{yang2021characterizing}, hardware heterogeneity~\cite{xu2021helios}, statistical heterogeneity, and label heterogeneity, lead to convergence problems of the global model~\cite{li2020federated}. In this paper, we focus on identifying data heterogeneity issues (i.e., statistical heterogeneity and label heterogeneity). Related studies leverage automatic solutions to improve model robustness. FedProx~\cite{li2018federated} modifies the weight updating algorithm to limit the impact of local updates on the global model. This approach, however, can hardly address the label heterogeneity problem, because the trained model can be meaningless if the definition of labels varies for different clients. Another solution is to train separated models instead of a global one. Separated models can learn from each other based on state-of-the-art frameworks, such as meta-learning~\cite{khodak2019adaptive} and multi-task learning~\cite{smith2017federated}. However, these automatic solutions lack the ability to analyze heterogeneity and therefore miss the chance to optimize models by means, like managing distributed data.

In the second scenario, meta-analysis~\cite{rosenthal2002meta} is proposed as a methodology to discriminate, combine and summarize multiple statistical analysis results. In the discrimination stage, statistical tests are leveraged to assess the significance of heterogeneity~\cite{higgins2002quantifying, langan2019comparison}. If the heterogeneity is significant, the statistical analysis results cannot be integrated directly~\cite{guler2018heterogeneity}. A feasible solution is to locate and exclude the variables or moderators that result in the heterogeneity~\cite{rosenthal2002meta}. However, these approaches are not practical for FL due to the privacy limitation. 

In this paper, we leverage limited available information (the parameters and the output of HFL models) to facilitate heterogeneity detection and examination, so that users can better understand the significance of heterogeneity issues.

\subsection{Visualization for Model Diagnosis}
Approaches for model diagnosis mainly fall into three categories: 
monitoring the model performance fluctuation, inspecting model configuration, and leveraging instance-level analysis. 

Visualization techniques aim to provide an overview of the model evolution. In existing applications, the performance of the model (e.g., loss) is frequently recorded and represented as time series~\cite{cashman2018rnnbow, jaunet2020drlviz}. We also show the performance dynamics in our system to assist users in judging the convergence and performance of the federated model.

To reason why the performance fluctuates, users need to further inspect the model configuration. DeepEyes~\cite{pezzotti2017deepeyes} allows users to check details of detected stable layers in deep neural networks through three linked views, which depict activations, instances, and filters to judge if a layer is oversized or unnecessary. GANViz~\cite{wang2018ganviz} compares image features from the dimension of time. The comparison results reflect the impact of a feature during the model training process. DGMTracker~\cite{liu2017analyzing} applies a credit assignment algorithm to locate the neurons that contribute to the failure~\cite{liu2017analyzing}. However, various models with distinctive structures can be trained based on the federated architecture. We focus on the HFL process, namely the exchange of parameters between the local client and the server, from which the disagreements between the local client and the others can be studied.

Testing output requires a lower learning cost than model interpretation, which is more acceptable for domain experts. To diagnose and improve the model, failed cases should be emphasized. The What-If-Tool~\cite{wexler2019if} allows users to customize the input of models and learn the mechanics of models by comparing related outputs. RetainVis~\cite{kwon2018retainvis} allows users to modify the input of an RNN model and figure out why a record is classified incorrectly. Krause et al.~\cite{krause2017workflow} provide instance-level explanations to verify the effectiveness of features based on a single instance. However, none of the above studies consider  data heterogeneity due to the difference between application scenarios.

\subsection{Visualization for Model Comparison}
Existing studies on model comparison mostly aim at selecting the best model, which requires to compare model performance. Instance-level analysis can be leveraged to facilitate the understanding of model behaviors. DeepCompare~\cite{murugesan2019deepcompare} groups instances by the combination of classification results of a Convolutional Neural Network (CNN) and a Long Short-Term Memory (LSTM) model. The group information indicates where the two models disagree with each other. Users can start exploration from the instances that are misclassified by one or two models. The neuron activation pattern of the user-selected instance can be compared through a heatmap. However, models can achieve the same label with a different confidence~\cite{ren2016squares}. Manifold~\cite{zhang2018manifold} employs scatterplots with color encodings to show the overview of instance-level details, which are the confidence of the model pair and the ground-truth label.

Model comparison can also offer an in-depth understanding of the data. On the one hand, the output of the model reflects the underlying characteristics of the data. Employing multiple models allows users to learn data from multiple perspectives. Alexander and Gleicher~\cite{alexander2015task} compare the results of two topic models to illustrate the consistency of documents. They introduce two visual encodings to represent the outputs of the two models, respectively. PK-clustering~\cite{pister2020integrating} compares the results of multiple clustering algorithms to reduce uncertainty in prior knowledge. The trend patterns can highlight inconsistencies. ConceptExplorer~\cite{wang2020conceptexplorer}
compares the performance fluctuation of the online learning models trained with different time-series datasets. Users can judge if those data have experienced consistent evolution based on the proposed drift-level index. Similarly, we attempt to comprehend the heterogeneity between the local data and the data distributed on the other clients by comparing the stand-alone training model with the HFL model.

\section{Approach Designs}

In this section, we first introduce the background of horizontal federated learning and describe the design requirements.

\subsection{Horizontal Federated Learning Architecture}
\label{sec:fed}
The typical architecture of horizontal federated learning (HFL) consists of a set of \textit{clients} who own local data and a \textit{server} which hosts the exchange of parameters among clients. At the beginning of the training, the model settings are unified for initialization. After that, cooperated training performs iteratively. Each communication round includes four steps. First, each client receives centralized parameters from the server. Second, each client updates the parameter by training and testing with local data. Then, each local update is submitted to the server. For privacy concerns, parameters can be encrypted~\cite{zhang2020batchcrypt} or processed by differential privacy~\cite{wei2020federated}. Third, the server integrates the parameters from the clients by aggregation algorithms. (We employ FedAvg~\cite{mcmahan2017communication} as default aggregation algorithms in this work.) Finally, the server sends the integrated results back, based on which each client updates the parameter of their own model and gets ready for the next round. In conclusion, each client can only access the parameters of the HFL model in the entire learning process.

 
\subsection{Requirements Analysis}
\label{sec:req}
Because clients cannot exchange the raw data, the primary issue of designing a visual diagnosis tool for data heterogeneity is that a direct comparison among data from different clients is infeasible. We interviewed two FL experts (E1 and E2) to define requirements. Both of them are senior machine learning researchers who have studied FL for more than 3 years and published related papers. They said that they have no effective solutions to deal with heterogeneity issues caused by heterogeneous data. Although a visualization system~\cite{li2021inspecting} is proposed to inspect the training process of the HFL, a tool for heterogeneity exploration is still missing. Without the understanding of data heterogeneity, it is challenging to remove heterogeneity issues thoroughly. Therefore, the experts agree that an interactive visualization tool to examine the data heterogeneity is critical for real applications. 

To identify the major requirements, questions are asked to two FL experts: Without the prior knowledge of data from other clients, how can you be aware of the existence of data heterogeneity? What kind of information is helpful in identifying and understanding the data heterogeneity?

Based on the interviews and literature review, we identify three requirements.

\textbf{R1. Knowing the existence of data heterogeneity.}
Both experts pointed out that it is critical to know the occurrence of data heterogeneity during the model building, so that the investigation of data heterogeneity can be performed at appropriate rounds.
\begin{itemize}
\item\textit{R1.1 Tracking the dynamic performance of the federated model.} 

Data heterogeneity will hinder the convergence of the federated model and harm its performance. 
The fluctuation of performance metrics, such as loss, is a good indicator of the training process~\cite{cashman2018rnnbow, jaunet2020drlviz}. 
Tracking them helps to know the dynamics of the training process and further locate the rounds when heterogeneity issues occur and affect the federated model.

\item\textit{R1.2 Inspecting the parameter conflicts between the client model and federated model.}
The federated model iteratively integrates model parameters from client models. The data heterogeneity will lead to differences in model parameters submitted by different clients. Therefore, the significant parameter conflicts are also a signal of possible data heterogeneity. Due to the number of model parameters, users need an efficient analysis manner to inspect the differences.
\end{itemize}
\textbf{R2. Comparing data in terms of model behaviors.}
Both experts agreed that the most critical barrier before investigating the data heterogeneity is the data isolation in FL. The only feedback information from other clients is the federated model. Instead of comparing data directly, E1 suggested that we can compare prediction behaviors of the models trained with the data. Because a model has learned from training data in the training process, based on which the model can make predictions for other inputs.

\begin{figure*} [!htbp]
   \centering
   \includegraphics[width=1.99\columnwidth]{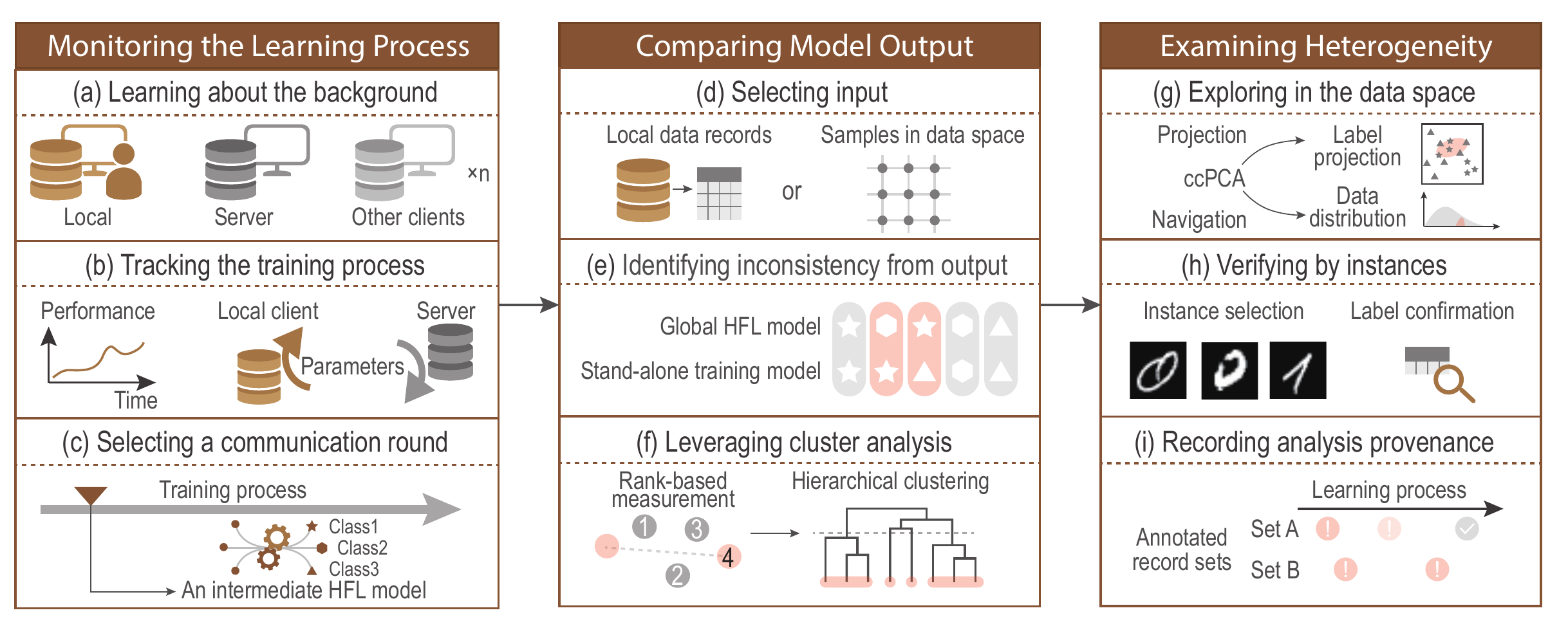}
   \caption{The three-stage workflow for analyzing heterogeneity issues in a federated cooperation from the perspective of a client.}
   \label{fig:wf}
\end{figure*}

\begin{itemize}


\item\textit{R2.1 Training a local model with the local data.}
The HFL model learns from global data during the training process. However, the client model learns more than the local data because the client model is updated with centralized parameters (see Sec~\ref{sec:fed}). We need a local model whose training process employs local data only. Besides, to exclude impacts of model designs, the local model should be trained in the same architecture (e.g., CNN models in the same network structure) as the HFL model. 



\item\textit{R2.2 Testing model output with various inputs.} 
Comparing model output is a common practice to learn the differences in prediction behaviors of two models~\cite{zhang2018manifold}. For a comprehensive comparison, we should provide inputs that can cover the corresponding data space. 

\end{itemize}  

\textbf{R3. Supporting visual examination of heterogeneity issues.} 
Visual analysis approaches can facilitate users in understanding heterogeneity issues and reasoning their impacts.
\begin{itemize}
\item\textit{R3.1 Identify heterogeneity issues based on inconsistent outputs.} 
E2 said that clustering records with inconsistent model outputs can facilitate in identification of heterogeneity issues. According to the theories in meta-analysis~\cite{rosenthal2002meta}, researchers need to deal with heterogeneity issues based on their significance. After clustering records, the significance of heterogeneity can be observed from the size of each cluster. If the ground-truth label is available, we can further calculate the prediction accuracy of the HFL model for the records in each cluster and assist client analysts to judge whether the impact of the corresponding heterogeneity issue is positive or negative.


\item\textit{R3.2: Reason the impact of heterogeneity issues by individual records.} 
Overview of a group can facilitate analysts to come up with hypotheses. A hypothesis could be ``two models strongly disagree in the classification of records with certain features.'' Analysts can realize significant data heterogeneity and infer that the federated model may yield wrong judgments for records with these features, according to the verified hypothesis. To make verification, the experts would like an instance-level visualization to access the details, such as the images and labels.


\item\textit{R3.3 Tracking the identified issues during the training process.} During the training process, the classification results for certain records may update. An intermediate result of a certain communication round may be randomly influenced (e.g., the training data included outliers accidentally), even after the model has converged. To draw a firm conclusion, analysts need to collect suspected issues and track them in the following communication rounds. Besides, analysts can observe what problems HFL faces and whether it can solve them by tracking the training process. It is significant to evaluate the effectiveness and robustness of the HFL model. 
\end{itemize}

\section{Workflow}
To support users, i.e., client analysts, analyzing heterogeneity issues, we propose a three-stage workflow (see Figure~\ref{fig:wf}): 1) monitoring the HFL process, 2) comparing model output, and 3) examining heterogeneity.

\subsection{Monitoring the Learning Process}
In the first stage, users need to learn the model configurations of the global HFL model and observe the training process. Model configurations include the choice of aggregation algorithms, the number of clients, and the description of the local data. The training process of HFL can be described by performance fluctuation (\textit{R1.1}) and the exchanges of model parameters between the local client and the server (\textit{R1.2}, Figure~\ref{fig:wf}(b)). Dramatic performance fluctuations or continuous conflicts of model parameters may indicate unsatisfying global data, which can motivate in-depth analysis for heterogeneity issues. The following analysis needs to be implemented based on a static model. To specify the intermediate HFL model, users need to select a communication round according to the observation results of the training process (Figure~\ref{fig:wf}(c)).

\subsection{Comparing Model Output}
\label{sec:st2}

In the second stage, users need to indirectly compare global data and the local data by comparing prediction behaviors of the two models trained  respectively. We employ local data to train a stand-along training model in the same model architecture with the global HFL model (\textit{R2.1}). The stand-along training model is used to compare with the HFL model, which is trained with global data. To collect various outputs, we provide three sets of data records as input (\textit{R2.2}, Figure~\ref{fig:wf}(d)), which are local data and two automatically generated datasets (please refer to Section~\ref{sec:sample} for details). The records with inconsistent outputs (records classified as different labels by the two models), or we say in the following, \textit{inconsistent records}, are then identified (Figure~\ref{fig:wf}(e)). Users can leverage a cluster analysis method (\textit{R3.1}, Figure~\ref{fig:wf}(f), Section~\ref{sec:cluster}) to generate \textit{inconsistency clusters} from inconsistent records.

\subsection{Examining Heterogeneity}
In the third stage, users examine an inconsistency cluster to study the corresponding heterogeneity issue and understand the impacts of the heterogeneity issue. As shown in Figure~\ref{fig:wf}(g), users can check data space characteristics of the records in the selected inconsistency cluster. Related findings inspire users to come up with hypotheses on suspicious heterogeneity. To make tentative verification, users can select certain records in the inconsistency cluster and browse record details (Figure~\ref{fig:wf}(h)). Records with ground-truth labels can facilitate users to judge whether the heterogeneity issue leads to a higher accuracy than the stand-alone training model (\textit{R3.2}).

As shown in Figure~\ref{fig:wf}(i), users can annotate the records with suspicious heterogeneity issues and track these records in the following training process (\textit{R3.3}). If the negative impacts of these records is weakened or even disappears, users can consider that the HFL models can overcome such heterogeneity. If not, this indicates users should  pay attention to possible heterogeneity issues, and reassess the cooperation of FL.



\section{Models}
In this section, we introduce the two models employed in our system.

\subsection{Input Generation from the Data Space}
\label{sec:sample}

We generate a representative input that distributes all over the data space for a comprehensive test of output comparison. Because the original data  space is often  of too high dimensionality to perform an efficient sampling, we assume that the data is distributed in a low-dimensional subspace. Therefore, we perform PCA on the data to get the low-dimensional space. After that, we employ stratified sampling, which is more efficient and effective than random sampling, to sample inputs in the low-dimensional space.

The low-dimensional samples are then projected back to high-dimensional space to satisfy the input format. To avoid illegal input, range interception is implemented according to the definition of each dimension. For instance, in a handwritten digit dataset, the grayscale of a pixel is regarded as a dimension whose threshold is from 0 to 255. If a dimension is restored as 260, we correct it as 255 for legality and validity. 

\subsection{Context-aware Clustering Approach}
\label{sec:cluster}
It is time-consuming to browse each inconsistent record from the high-dimensional space to compile a   summary on heterogeneity issues. We, therefore, leverage a cluster analysis to organize inconsistent records and analyze heterogeneity. 
Observing clustering results with different settings of the cluster number can facilitate users to generalize heterogeneity issues from different levels. To support flexible adjustment of the cluster number, we employ hierarchical clustering to group inconsistent records. 

However, inconsistent clusters may be mixed up with consistent records, which hardly contribute to extraction of heterogeneity issues. To exclude consistent records from inconsistency clusters, we have to consider the context of consistent records surrounding inconsistent records when constructing hierarchical clustering. Therefore, we change the distance measurement in the hierarchical clustering from Euclidean measurement to a rank-based measurement. We denote the set of all records as $\{s_i|i = 1,\dots, n\}$. Among them, there exist $m$ inconsistent records, which are $\{s_i|i = 1,\dots, m\}$ ($m\leq n$). The Euclidean distance from $s_j$ to $s_k$ is represented by $d_E(j,k)$. The rank-based measure calculates the distance between a pair of inconsistent records $s_j$ and $s_k$ ($1\leq j\leq m$, $1\leq k\leq m$, and $j\neq k$) as:
\begin{equation}
d_R(j, k) = r_j(k)\times r_k(j),
\end{equation}
where $r_j(k)$ is the ranking of $d_E(j,k)$ in $\{d_E(j,i)|i = 1,\dots,n\}$. Namely, $s_k$ is the $r_j(k)$-th closest record to $s_j$ among all records, including consistent records.
Based on the rank-based measurement, inconsistent records with less surrounding consistent records will be preferentially aggregated together, even if the inconsistent records are far apart.

\section{Visual Analysis Interface}
Corresponding to the workflow, the interface of \techname consists of three modules, which are learning process monitor (Figure~\ref{fig:tea}(a)), output comparison (Figure~\ref{fig:tea}(b)), and heterogeneity examination (Figure~\ref{fig:tea} (c)).

\subsection{Learning Process Monitor}
The module for monitoring the HFL process (Figure~\ref{fig:tea}(a)) consists of an information panel, a parameter projection view (Figure~\ref{fig:tea}(a1)), and a performance view (Figure~\ref{fig:tea}(a2)). The information panel, which introduces the configuration information of the HFL model and the description of the analyzed client, are listed at the top of the module. The descriptions will be updated with the progress of the model training (i.e., a new communication round is finished).

\textbf{Parameter projection view: }
As shown in Figure~\ref{fig:tea}(a1), the parameter projection view summarizes exchanges of model parameters between the local client and the server. Because model parameters updated in a single communication round could be affected by accident outliers, it is necessary to provide an overview of parameter updates during the entire training process. In each communication round, model parameters can be considered as a high-dimensional vector. To observe parameters from a low-dimensional perspective, we generate a 2D projection for the parameter vectors of the HFL model in all communication rounds. To reflect parameter exchanges, parameters submitted by the local client are transformed into vectors of the same size and projected onto the same plane. The employed projection approach is accelerated by a probabilistic algorithm~\cite{halko2011finding}. Considering that different parameters contribute in different ways, we allow users to specify a group of parameters and check them individually. For example, users can specify a layer in the neural network and check the projection of the model parameters on this layer. 

In the parameter projection view, the points that project model parameters of the HFL model in each communication round are connected by a gray polyline. The grayscale encodes the time sequence. A brown arrow is drawn from the point projecting federated parameters in round $i-1$ to the local client parameters in round $i$. The federated parameters in round $i$ are different from the local parameters because the federated parameters have integrated parameters from other clients. The size of the angle between the arrow and the polyline implies the disagreement of parameters at round $i$. The cosine of each angle in the high-dimensional space is calculated and encoded by the darkness of the corresponding arrow to avoid misunderstanding caused by projection distortion. Disagreements and compromises between the local client and the server can be observed by comparing historical parameter updates with the actual parameter evolution of the HFL model. 

\textbf{Performance view:} At the bottom of the interface, the performance view (see Figure~\ref{fig:tea}(a2)) monitors the dynamics of the performance testified by the local data. Performance indicators consist of the training loss, the accuracy for the test set, and the total accuracy for local data. Users are allowed to switch indicators to apply a comprehensive evaluation. 

\subsection{Output Comparison}
The second module (Figure~\ref{fig:tea}(b)) consists of an output comparison view and a summary of inconsistency clusters. 

The result of output comparison is displayed in the output comparison view (Figure~\ref{fig:tea}(b1)). To contrast inconsistent records (in brown) with others (in gray), all records are projected through ccPCA. As mentioned in Section~\ref{sec:st2}, the projection could hardly split all inconsistent records from the rest simultaneously. Users are allowed to check the overlaps by switching the top layer between inconsistency and consistency. 

We list inconsistency clusters in the order of size  (Figure~\ref{fig:tea}(b2)) to motivate heterogeneity examination. Each inconsistency cluster is represented by a glyph. In each glyph, the convex hull of the inconsistent records is superimposed on a density heatmap of all records in the output comparison view. The cluster size and the accuracy of the HFL model on the cluster are listed below the corresponding glyph. A cluster with an extremely low accuracy may suggest a significant impact from the corresponding heterogeneity issue. Clusters with insufficient records can be regarded as outliers. 

\subsection{Heterogeneity Examination}
Users examine an inconsistency cluster in the third module of the interface (Figure~\ref{fig:tea}(c)), which includes two views for distribution exploration, an instance verification panel (Figure~\ref{fig:tea}(c4)), and an annotations panel (Figure~\ref{fig:tea}(c5)). The distribution exploration views are introduced as follows.

\textbf{Dimension exploration view:}
Users need to extract heterogeneity issues by identifying commons shared by inconsistent records but not shared by others. However, it is exhausting to inspect each dimension, considering the records are high dimensional. To improve analysis efficiency, we provide users with two entrances for dimension selections. The first entrance navigates users from the perspective of data space. This entrance shows how important each dimension is to distinguish the selected cluster from others based on the first two cPCs of ccPCA~\cite{fujiwara2019supporting}. The second entrance navigates users from the perspective of model behaviors, which only works for HFL based on CNN models. Gradient class activation maps (Grad-CAM)~\cite{Selvaraju2017grad} can measure how important each dimension is for a CNN model to classify a record. The average Grad-CAM of all records in the cluster can identify significant dimensions to a model. To analyze inconsistent model behaviors, we generate a pair of average Grad-CAMs for the stand-alone training model and the HFL model, respectively. Comparing the two average Grad-CAMs can help users identify the differences in model judgments. However, such differences in model judgment can hardly be demonstrated by Grad-CAMs in rare cases (e.g., models with cascading randomization)~\cite{Adebayo2018, kindermans2019reliability}. It is necessary to leverage both of the entrances.

We employ a pair of pixel maps to represent the user-specified dimension entrance. Note that both entrances generate a pair of importance values. When the entrance based on ccPCA is activated, two pixel maps represent the quantified importance of each dimension measured by the first two cPCs, respectively. When the entrance based on Grad-CAM is used, the two pixel maps represent the quantified importance evaluated by the two models, respectively. In a pixel map, each pixel corresponds to a dimension. If records are in the form of pictures, the relative position of a pixel is consistent with its position in pictures. If the pictures have more than one channel (e.g., RGB channels), users can select a channel and check a channel at a time. The color of a pixel encodes the dimension importance.

Users can hover on a pixel to find the pixel corresponding to the same dimension in the other pixel map. Users can also click a pixel and check the corresponding dimension distribution of all records or the records in the cluster (see Figure~\ref{fig:tea}(c2)). Percentage distributions of overall records, inconsistent records, and consistent records are displayed respectively. To adapt to different distribution patterns, the scale on the y-axis (percentage) can be switched from linear to logarithmic.

\begin{figure} [!htbp]
  \centering
  \includegraphics[width=0.9\columnwidth]{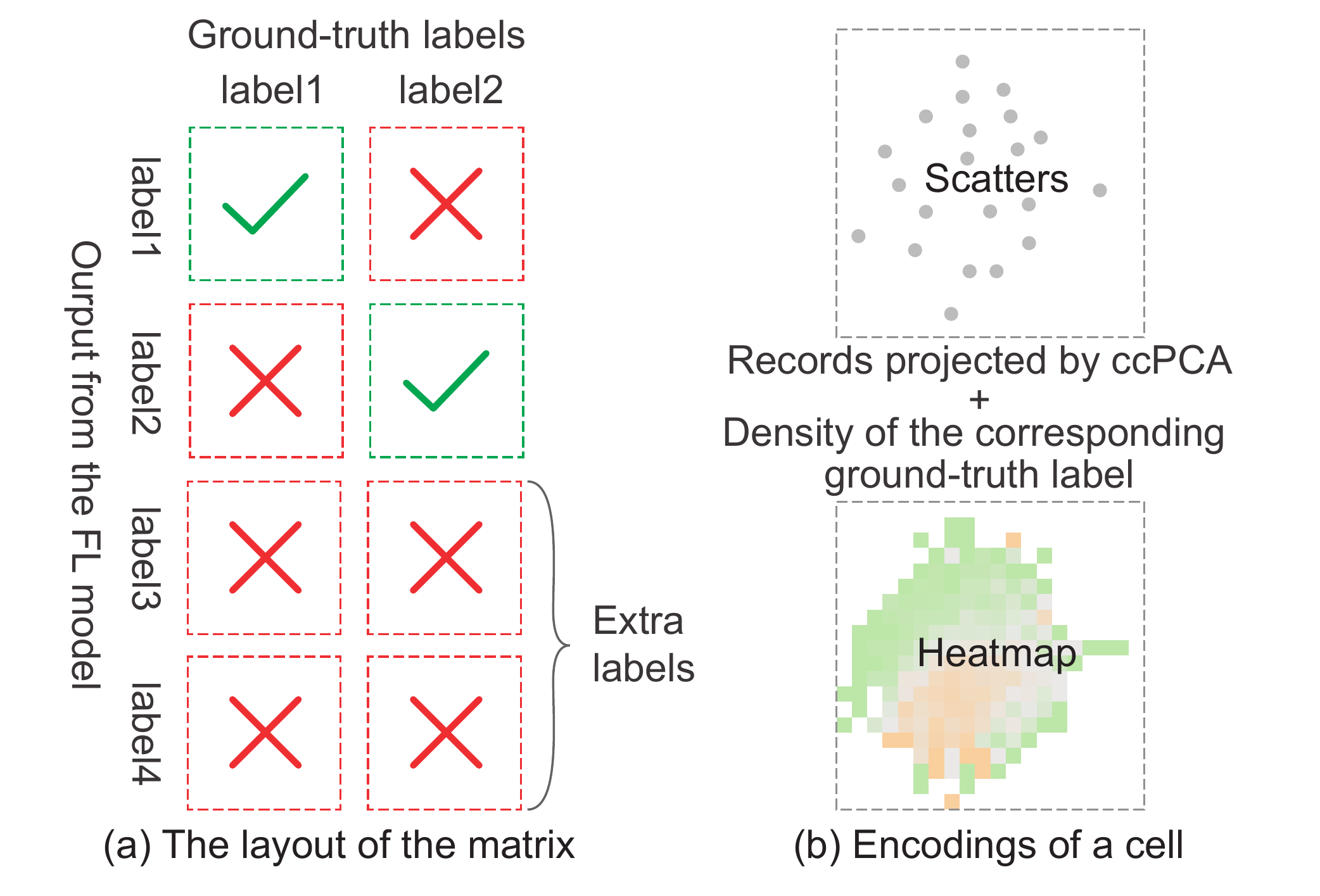}
  \caption{The visual design of the label exploration view.}
  \label{fig:lab}
\end{figure}

\textbf{Label exploration view: }
A matrix design is leveraged to compare the ground-truth labels with the output of the HFL model (see Figure~\ref{fig:tea}(c3)). Each cell of the matrix corresponds to a pair of labels (see Figure~\ref{fig:lab}(a)). The horizontal position is specified by the ground-truth label, and the vertical position is specified by the output from the HFL model. The records in the non-diagonal cells indicate that the HFL model generates incorrect output. If extra labels (i.e., the labels not included in the ground-truth labels) exist in the output of the HFL model, there will be extra rows listed after the labels included in the ground-truth label. Users can scroll down to check the cells corresponding to extra labels. 

The records, which meet the pair of the ground-truth label and the output of the HFL model, are projected in the corresponding cell as a scatter plot by ccPCA~\cite{fujiwara2019supporting} (see Figure~\ref{fig:lab}(b)). For each cell, we count the inconsistent records in the selected cluster and the local data, respectively. The numbers are displayed in the upper-left corner (see Figure~\ref{fig:tea}(c3)). The inconsistent records and the consistent records are with the same color encoding as those in the output comparison view. Also, the convex hull of the selected cluster is drawn in each cell. 

A grid-based heatmap is shown as a background in each cell (see Figure~\ref{fig:lab}(b)). The background color encodes the density of the corresponding ground-truth label. Through comparing the background with scatters or convex hulls, users can come up with conjectures, like ``there exist label heterogeneity.'' Therefore, the cells in the same column are with the same background. The grid size can be adjusted to observe from different levels of granularity. Users are also allowed to hide the scatters and focus on the label distribution in cells. The contents in all cells can be zoomed in together.

\subsection{User Interactions}
Our system supports the following interactions.

\textbf{Check an intermediate result of the HFL model.} In the performance view, users can drag a handle to select a communication round and analyze the HFL model updated at this round. If the round is included in the user-specified range for the updates projection view, a circle appears in the projection view to highlight the corresponding round.

\textbf{Request for recommended parameters.} In consideration of efficiency, the initial contrastive parameters for ccPCA and the number of inconsistency clusters are default as 10 and 8. To seek better effectiveness, users can request recommended parameters by clicking the button on the right of the input box. The recommended contrastive parameter for ccPCA is provided by the original approach~\cite{fujiwara2019supporting}. The cluster number is recommended through the maximum difference~\cite{thorndike1953belongs}. 

\textbf{Examine an inconsistency cluster.} After browsing the list of inconsistency clusters, users can click on a glyph in the column of the model output comparison to examine it in the third module. The selected glyph will be highlighted by a thick stroke.

\textbf{Apply instance verification.} Users can select a record in the label exploration view by clicking. The record description, including dimension details, the ground-truth label, and the output of the two models, can be found in the instance verification view (see Figure~\ref{fig:tea}(c4)).

\textbf{Manage records with suspicious heterogeneity issues.} Users are allowed to select all records in the currently analyzed cluster as the object in the control panel  (see Figure~\ref{fig:tea}(c5)). Note that local data records are selected by the convex hull if the data records in the cluster are sampled data. If necessary, the intersection set or the joint set of the current cluster and annotated records can be selected. 

\textbf{Leverage the analysis provenance.} In the column of heterogeneity examination, users can annotate their findings and record the analyzed inconsistency cluster to the analysis provenance in the control panel (see Figure~\ref{fig:tea}(c5)). Each annotation generates a message icon in the performance monitor view. Users can click a message icon to review the details of previous annotations. Also, the records in the annotated set will be highlighted in both the model output overview and the label exploration view. If the annotated cluster has an overlap with an inconsistency cluster, the cluster glyph will be highlighted, and the size of the  overlap will be shown below the glyph. Each annotation can be deleted by right-clicking on the icon.

\section{Case Studies}
\subsection{Handwritten Digits Recognition}
The MNIST dataset~\cite{mnist} is employed in the first case to train a Multilayer Perceptron (MLP) for classification. The original dataset was distributed to 10 clients. Each client has 6,000 records of pictures, with two consecutive digital labels, e.g., digit-1 and digit-2. $10\%$ records are employed in the test set. The analyzed client owns records with labels consisting of digit-0 and digit-1.

\begin{figure} [h]
   \centering
   \includegraphics[width=0.99\columnwidth]{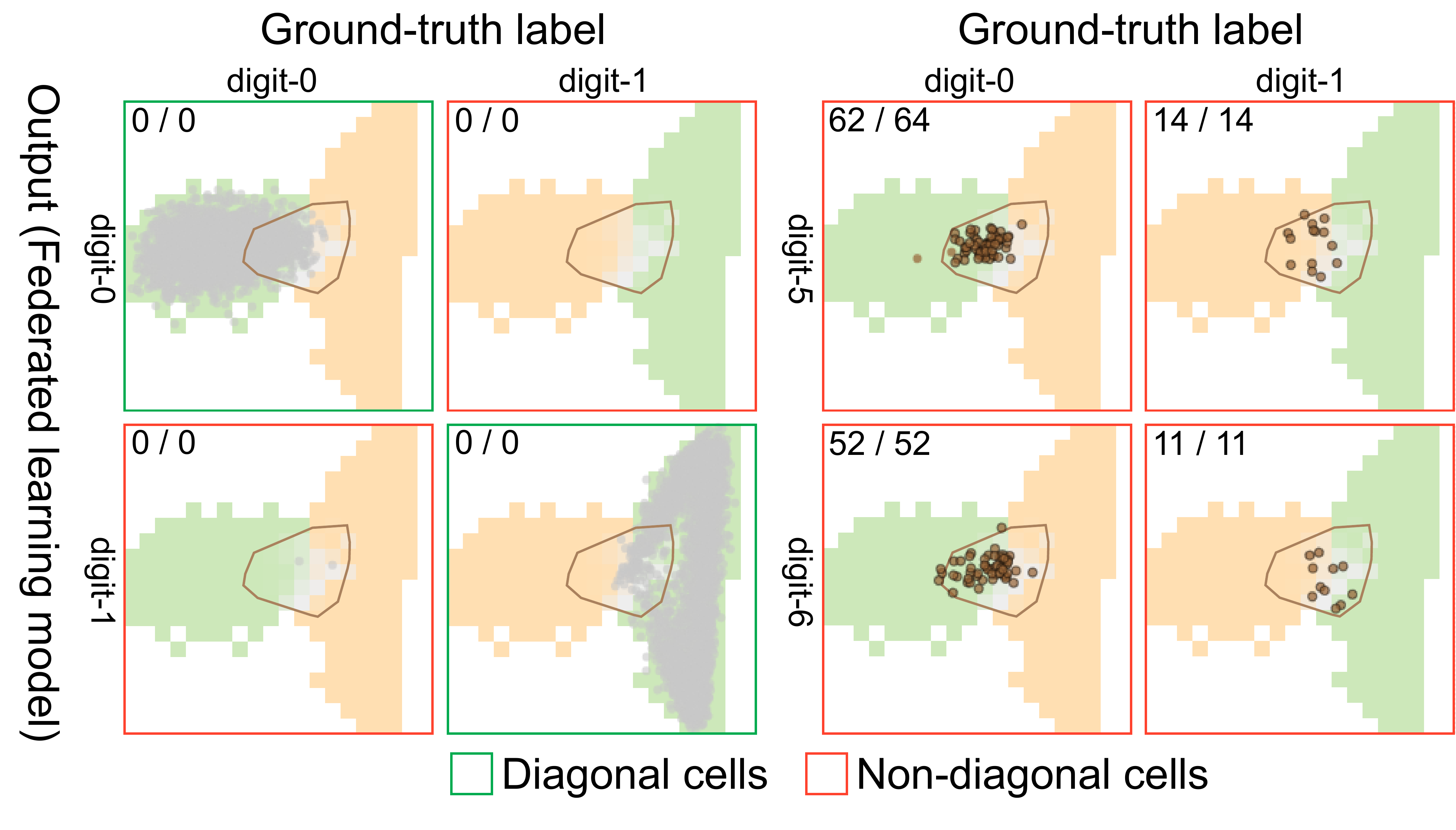}
   \caption{Partial matrices in the label exploration view. A majority of inconsistent records distributed in the cells whose output from the HFL model includes other labels, like digit-5 and digit-6.}
   \label{fig:c1m}
\end{figure}

As shown in Figure~\ref{fig:tea}(a2), the loss became stable at the 40th round (\textit{R1.1}). We check the convergence process by selecting this round. We employed local data as input to compare outputs from the stand-alone model and the HFL model (\textit{R2.2}). 300 inconsistent records were identified from the comparison. As shown in Figure~\ref{fig:c1m}, the numbers of inconsistent records in the label exploration matrix reflect that most inconsistent records locate in non-diagonal cells, where ground-truth labels are different from the output of the HFL model. The HFL model has a lower accuracy than the stand-alone training model. The HFL model classified partial records as other labels.

\begin{figure} [h]
   \centering
   \includegraphics[width=0.99\columnwidth]{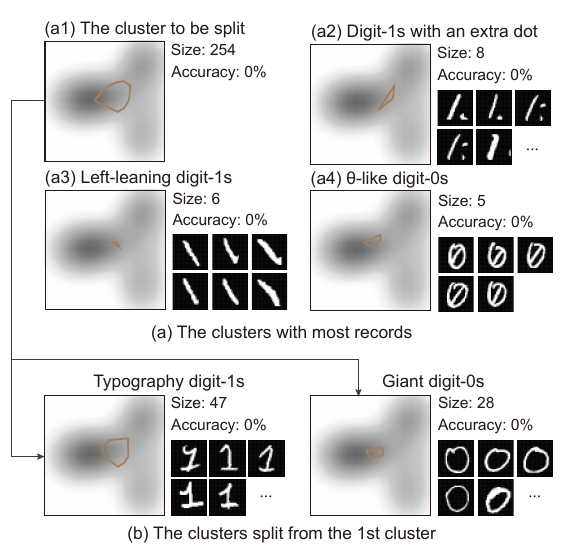}
   \caption{Inconsistency clusters of inconsistent records identified from the 40th round. (a) The top four clusters with the recommended cluster number. (b) Two clusters split from the 1st cluster in (a) after cluster number reaches 50.}
   \label{fig:c1c}
\end{figure}

We generated a ccPCA projection based on the recommended contrastive parameter to have an overview of all records. The projection result indicated that inconsistent records are not gathered closely and heavily overlapped with consistent ones. To distinguish them, we captured the features of inconsistent records by inspecting their clusters (\textit{R3.1}).
We paid attention to the four clusters with the most records. The second, the third and the fourth clusters correspond to a rare handwritten style, respectively (see Figure~\ref{fig:c1c}(a)). 

\begin{figure} [ht]
   \centering
   \includegraphics[width=0.99\columnwidth]{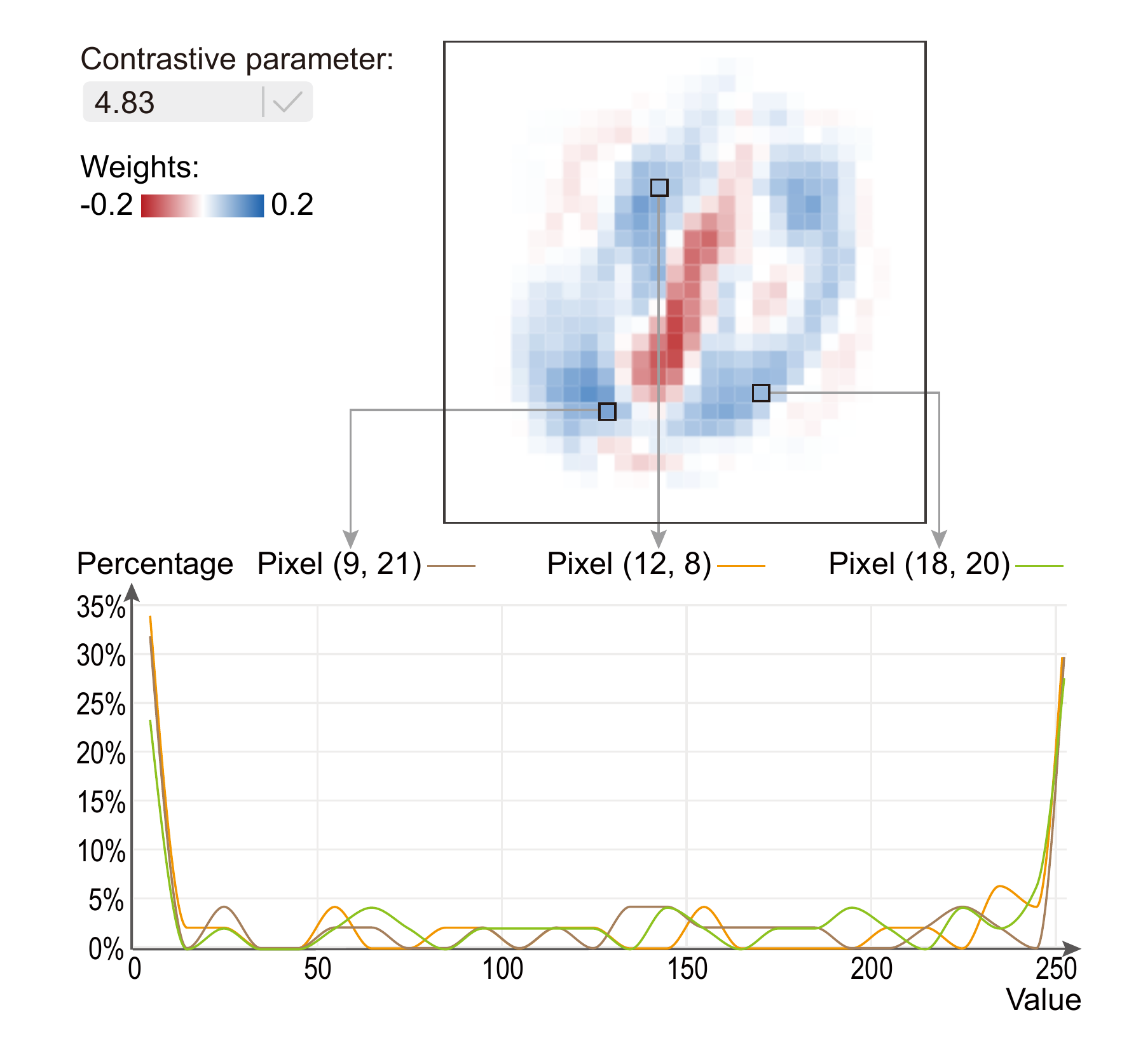}
   \caption{The distributions of three dimensions selected from the dimension weight chart. The high dimension values demonstrate that the corresponding pixel is passed by strokes.}
   \label{fig:c1w}
\end{figure}

We noticed that records with various handwritten styles are mixed up in the first cluster. To categorize styles and refine heterogeneity issues, we further split the first cluster up by setting a larger cluster number (see Figure~\ref{fig:c1c}(b)). One of them, consisting of 47 records, can not be further split up with a minor increase in the cluster number. The label exploration view indicates that all of the records are with a ground-truth label of digit-1. To explore this cluster, we checked the cluster distributions of several dimensions with a high weight assigned by ccPCA. As shown in Figure~\ref{fig:c1w}, certain dimensions are distributed in corners of the pictures. Verified by several records, we found that they were digit-1s in typography font (\textit{R3.2}). The HFL model is inclined to 
misidentify this rare handwritten style as other digit labels, like digit-7. 

We annotated the 47 records in the cluster and inspected these records in the final round (i.e., the 200th round) (\textit{R3.3}). Partial records of digit-1 with an extra dot or giant digit-0 are corrected, while the rest are not (see Figure~\ref{fig:tea}(b2) and (c4) to inspect the cluster of digit-1s in typography font). Although the number of inconsistent records decreases to 208, the accuracy of the HFL model is still not satisfying as the stand-lone training model. 
Therefore, we should suggest other clients collect more pictures of digit-0 and digit-1 to optimize the HFL model.

\subsection{Face Mask Recognition}
In the second case, we trained a federated CNN model to recognize if the person in a color picture wears a mask. Two clients participated in the FL cooperation. Our client used the face mask image dataset provided by Jangra~\cite{fm1}. The other client employed the face mask detection dataset provided by Gurav~\cite{fm2}. All records are unified to RGB images in the size of 28*28 pixels. The local training set has 6,000 records and the local test set has 1,792 records.

\begin{figure} [ht]
   \centering
   \includegraphics[width=0.99\columnwidth]{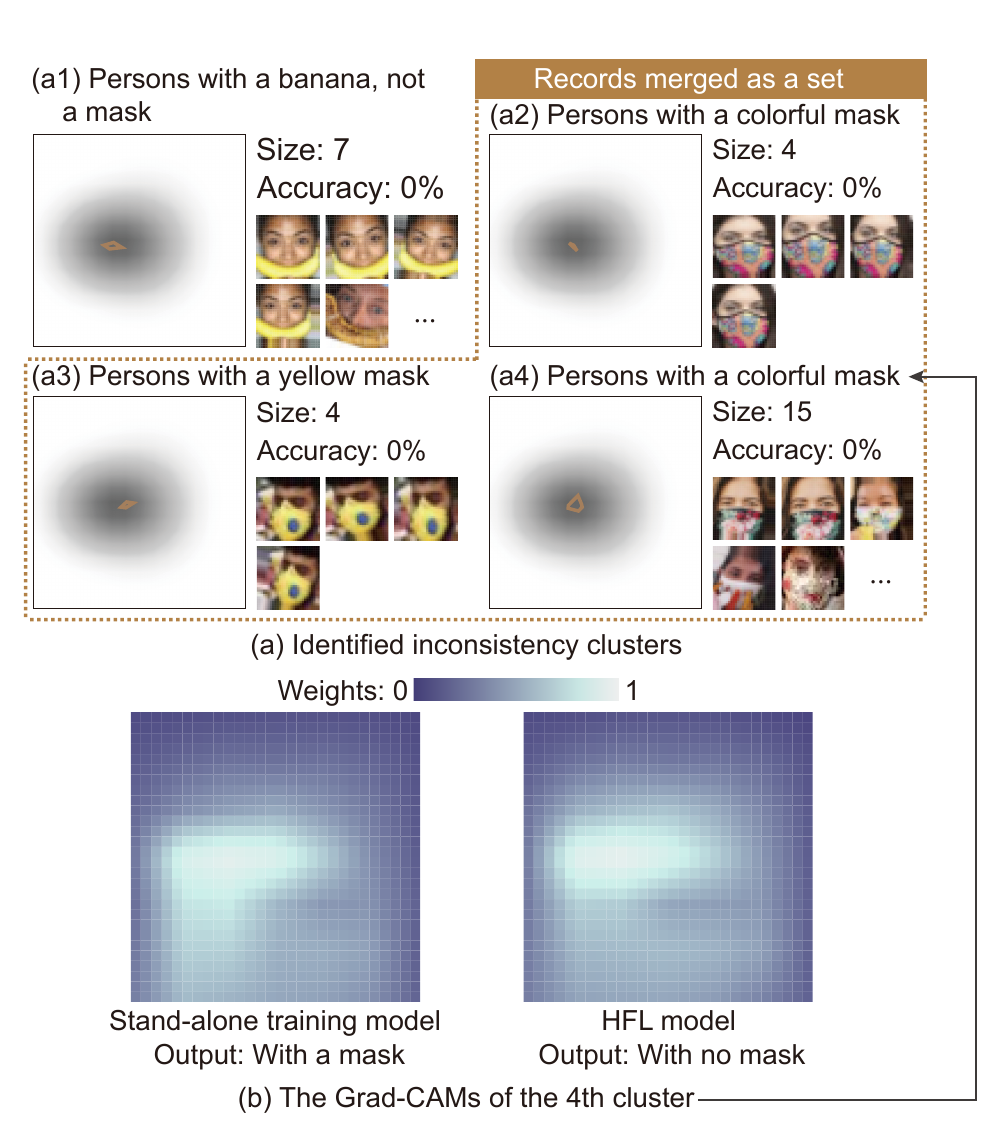}
   \caption{Cluster analysis of the inconsistency identified at the 90th round. (a) Inconsistency clusters with an accuracy of 0\%. (b) The Grad-CAMs of the 4th cluster in (a).}
   \label{fig:c2i}
\end{figure}

We first observed the learning process from the parameter projection view (\textit{R1.2}). To compare high-level features extracted by CNN models, we selected the last two convolutional layers, respectively. Both the projected polylines of the parameters become relatively stable at the 90th round. We grouped the 547 inconsistent records identified at this round into 80 clusters (\textit{R3.1}). Three clusters with extra-low federated accuracy and a relatively large size caught our attention. 
All of the records in these clusters are misclassified as with no mask by the HFL model. We then checked these records by clicking the highlighted circles in the label exploration view. As shown in Figure~\ref{fig:c2i}(a1), the first cluster with 7 pictures corresponds to a person who puts a banana on the mouth to disguise a mask. These records are labeled incorrectly. In the other two clusters, the masks with special patterns are neglected by the HFL model (see Figure~\ref{fig:c2i}(a2-a3)). We found more masks with special patterns after we set the cluster number as the recommendation (i.e., 125). As shown in Figure~\ref{fig:c2i}(a4), a new cluster with 15 records is identified. Similarly, all records are misclassified by the HFL model.

To investigate the reasons, we compare discriminative regions of the HFL model with the stand-alone training model. Both Grad-CAMs highlight the same region (see Figure~\ref{fig:c2i}(b)), which implies that the two models consider the same area as the basis of discrimination. However, the HFL model makes wrong judgments. We had a concern that the performance of the HFL model was affected because the data distributed in the other client is lacking our patterns. 

\begin{figure} [!htbp]
  \centering
  \includegraphics[width=0.99\columnwidth]{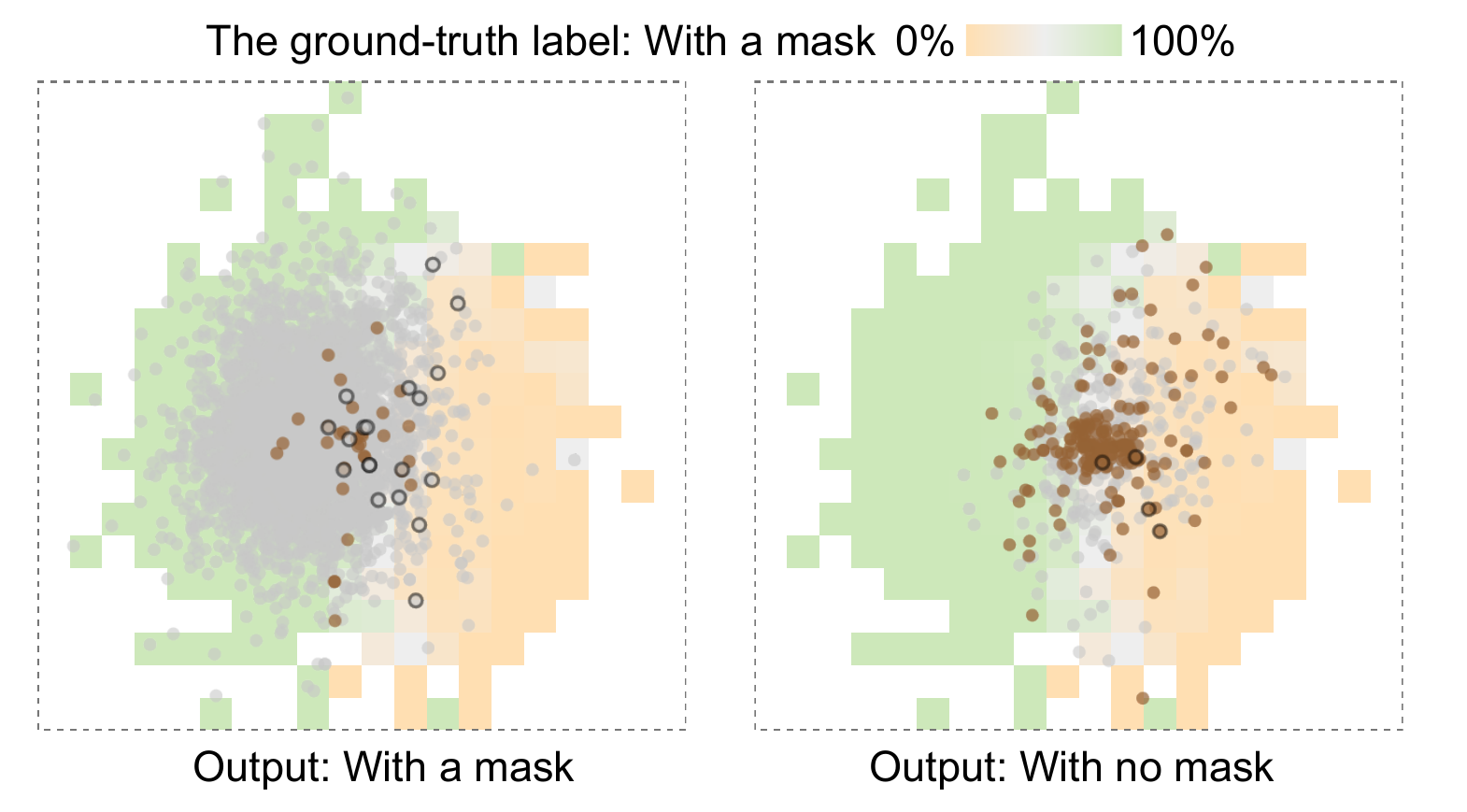}
  \caption{The label exploration view highlights the annotated records of colorful masks. Most of the records are consistent with the output of the stand-alone learning model, as well as the ground-truth labels.}
  \label{fig:c2m}
\end{figure}

To investigate the banana issue and the pattern issue, we annotated corresponding clusters and track them in the HFL process (\textit{R3.3}). In the 200th round, the total of inconsistent records has decreased to 347. We reviewed the annotated records. Most pictures of masks with patterns are classified correctly by the HFL model (see Figure~\ref{fig:c2m}). It turned out that the HFL model is capable of handling such statistical heterogeneity issues. However, a record with a ``banana mask'' was classified as with a mask. To optimize performance, we have to correct the labels of ``banana masks'' and notify the other client of this accident. 
We further compared model output by sample input (\textit{R2.2}). The convex hulls of the most inconsistency clusters have no overlap with local data records. To judge whether corresponding data heterogeneity exerts positive impact on our classification task, we need to collect corresponding records from real world and verify the output with ground-truth labels. 

\subsection{Vehicle Recognition}
In the third case, four clients seek accurate vehicle recognition by federated cooperation. Each client collected an equal number of pictures in three of four categories (i.e., plane, car, ship, and truck) from CIFAR-10 dataset~\cite{krizhevsky2009learning}. No picture is shared by two clients. The analyzed client owns 5,400 pictures: 1,800 pictures of plane, car, and truck, respectively. 900 pictures are included in the test set. 

\begin{figure} [!ht]
   \centering
   \includegraphics[width=0.99\columnwidth]{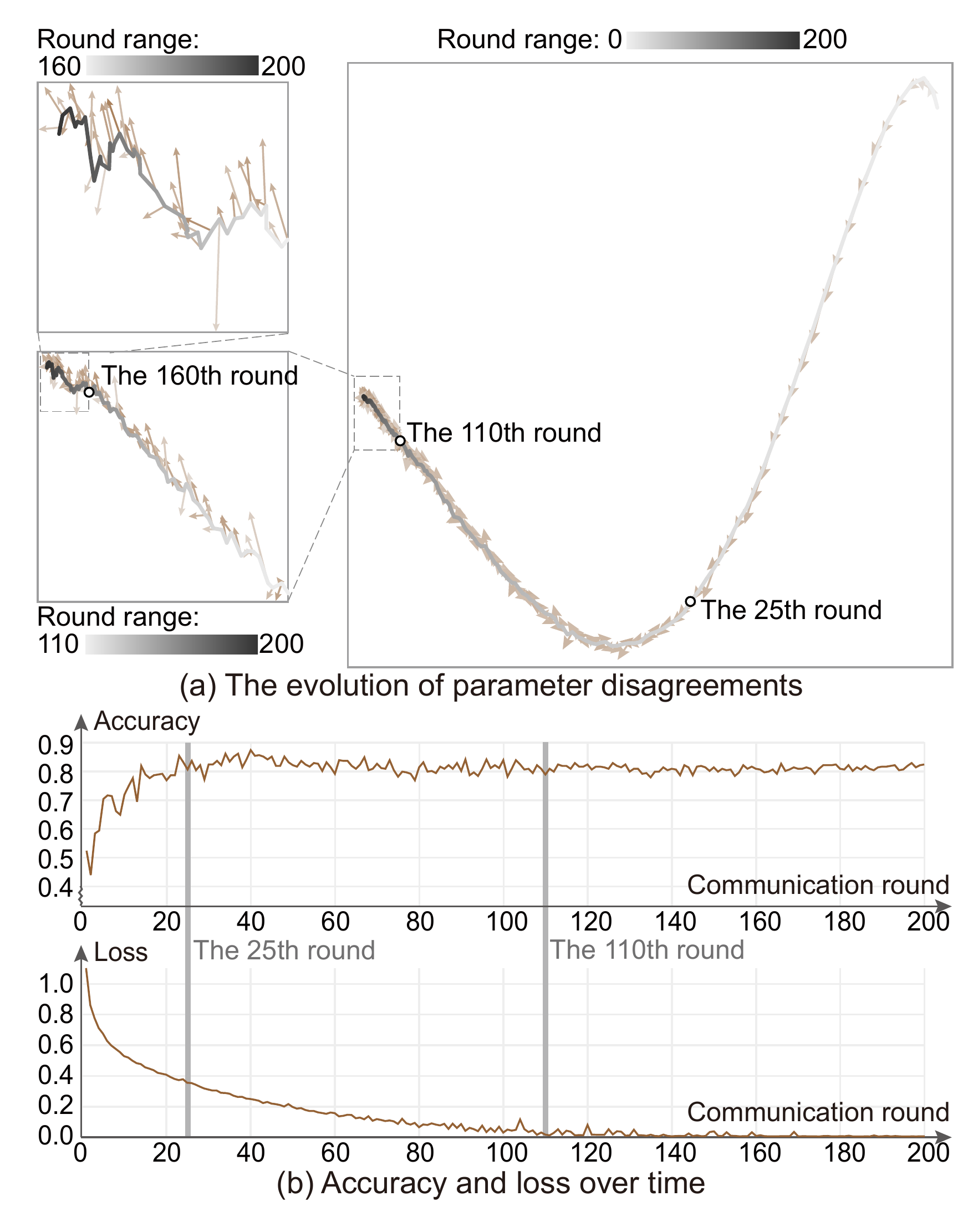}
   \caption{The training process of the HFL model. (a) The updates of the parameters in the last convolutional layer become unstable with the increase of the communication rounds. (b) The accuracy and the loss fluctuation during the training process.}
   \label{fig:c3o}
\end{figure}

According to the pattern of parameter fluctuation (\textit{R1.1}), we split the carried-out training process into three stages at the 25th round and the 110th round. As shown in the parameter projection view (Figure~\ref{fig:c3o}(a)), parameters of the HFL model change significantly in the first 25 rounds, which indicates a fast learning process (\textit{R1.2}). Then, the accuracy has approached its maximum around the 25th round (see Figure~\ref{fig:c3o}(b)). At the second stage, divergence could be observed between local updates and the HFL model from the directions of the arrows. At the end of the second stage (i.e., the 110th round), the loss has reached its minimum (see Figure~\ref{fig:c3o}(b)). In the last stage, the polyline in the parameter projection view fluctuates dramatically within a small range.  

We checked the end of each stage to investigate the evolution of heterogeneity issues. Among 955 inconsistent records at the 25th round, 90\% of them are classified incorrectly by the HFL model. The statistical results in the label exploration view demonstrate that the HFL model misclassified 425 records as ships (\textit{R3.2}). Besides, the HFL model encountered difficulties in distinguishing between cars and trucks, which leads to other 225 incorrect records.

\begin{figure} [!ht]
   \centering
   \includegraphics[width=0.99\columnwidth]{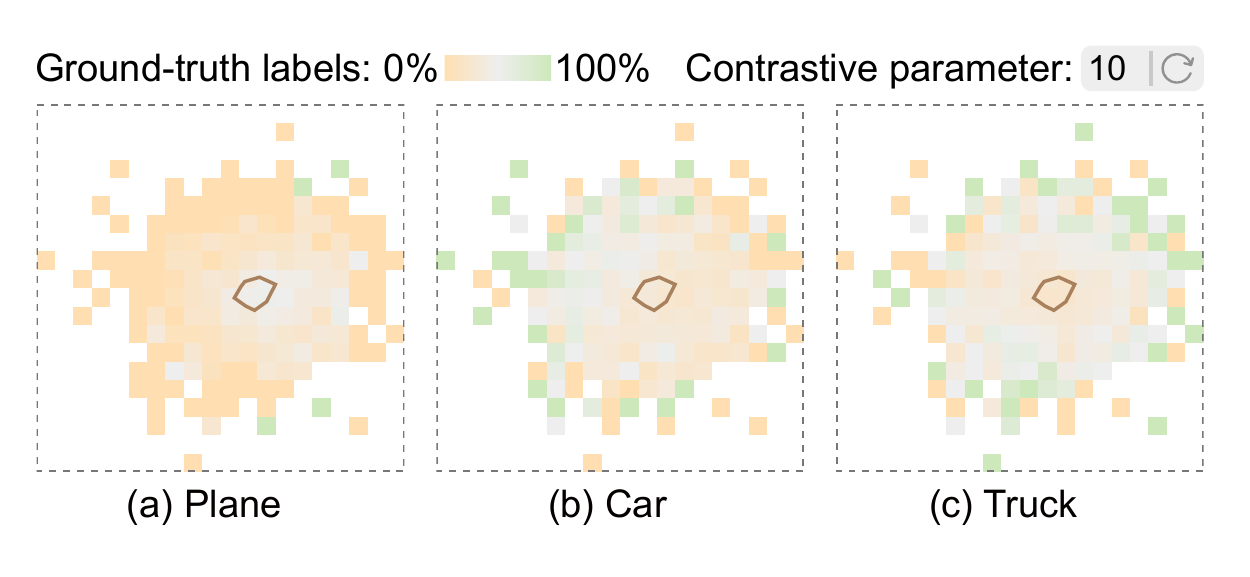}
   \caption{The distributions of the three labels.}
   \label{fig:c3m}
\end{figure}

In the 200th round, 162 records were still classified as a ship  (\textit{R3.3}). To focus on the label distribution, we hid the scatters. As shown in Figure~\ref{fig:c3m}(a), most planes were projected inside or beside the convex hull of inconsistent records, because planes have similar features to ships (e.g., with blue background). Meanwhile, records with labels of car or truck are distributed far from the convex hulls (see the green grids in Figures~\ref{fig:c3m}(b) and (c)). The HFL model is inclined to classify a plane as a ship.

Nevertheless, the car-and-track issue was drastically alleviated since the 110th round. To inspect the confusion among ground-truth labels at the final round, we checked the first three rows of the label exploration matrix, which corresponds to the records classified as plane, car, and truck by the HFL model. There are 36 inconsistent records in the non-diagonal cells (misclassified by the HFL model) and 157 ones in the diagonal cells (misclassified by the stand-alone training model). Therefore, the HFL model is superior to the stand-alone training model in the task of distinguishing local labels. In summary, we could benefit from the HFL cooperation. To fix the plane-ship issue, we should invite more partners to train the HFL model.
\section{Discussion}
In this section, we discuss 1) expert reviews on \techname, and 2) the result of a comparative study for the proposed context-aware clustering approach. 
\subsection{Expert Reviews}
We interviewed three researchers (E3, E4, and E5) who had worked on FL for two years. For each interview, we first introduced our approach and then presented a demo of our system. After that, experts were allowed to freely explore data heterogeneity of the above three cases in our system. At the end of the interview, we collected their feedback on the following four aspects. Each interview lasted about an hour.

\textbf{Effectiveness.} Due to data isolation, existing approaches to heterogeneity analysis are mainly by observing local data and identifying skewed distributions. All experts agreed that our system could  help them better to formulate reasonable hypotheses of heterogeneity than their original approaches. E3 said that the heterogeneity issues could be solved by adjusting records in the local client (e.g., expanding records) and findings in our system indicated which kind of adjustments could improve the performance of the HFL. E5 also noted that cluster analysis could efficiently locate heterogeneous issues and guide batch corrections.

\textbf{Usage experience.} The usability of our system received positive feedback from the experts. Both E3 and E4 appreciated the annotation functionality and commented that annotation allowed them to track certain sets of records. For visual designs, E3 and E5 were impressed by the intuitive representations in the parameter projection view. E3 commented that ``It can clearly reflect the conflicts between local updates and the HFL parameters.'' Particularly, a disagreement occurred with the dimension exploration. E4 was inclined to navigate by ccPCA because it could summarize data characteristics. However, E5 preferred the model-driven Grad-CAM to understand model behaviors.


\textbf{Findings.} 
When the accuracy of either model was satisfying, E4 found that the inconsistent records could depict the boundary of two classes in the ccPCA projection. Based on this observation, E4 drew the conclusion that the HFL model and the stand-alone training model would fail to reach an agreement especially when the target records were hard to classify, i.e., the records were distributed along the classification boundary. E4 also found that ccPCA separated inconsistent records from the consistent context, which in the meantime split different classes (see Figures~\ref{fig:tea}(b1) and \ref{fig:c2m}). This indicated that the projection results could also help users to assess model performance.

\textbf{Advice.} 
Despite the effectiveness and usefulness of our system, the experts offered two suggestions. First, we should take into account the architecture of the neural network when analyzing parameter exchanges (E3, E5). For example, users might focus on parameters in certain layers in the neural network when evaluating a deep learning model. Therefore, we improved the parameter projection view by grouping parameters based on layers. Users are allowed to select a layer and check parameters in the corresponding layer. Second, considering that dozens of labels might be yielded from the HFL model, E3 pointed out that the label exploration view needed an overview to facilitate object location. We plan to add it in the future version of our system.

\subsection{Comparative Study}
As mentioned in Section~\ref{sec:cluster}, we proposed a context-aware clustering approach to extract heterogeneity issues. 
To prove the effectiveness of our clustering approach, we compared our approach with the distance-based clustering approach based on the local data in the first case to seek instance-level verification. 

Two clustering approaches were applied to cluster the 208 inconsistent records identified in the 200th round. The maximum difference~\cite{thorndike1953belongs} was employed to recommend appropriate cluster numbers for both clustering approaches. Our clustering approach generated 59 clusters, among which the largest cluster consisted of 45 records of typographic digit-1s while the distance-based approach yielded 118 clusters, none of which contained more than 6 records. Although the records in the same cluster were similar to each other, similar records (e.g., typographic digit-1s) were split into different clusters.

To eliminate the effects of the recommendation algorithm, we set the cluster number to  100 for both approaches in the second experiment. The distance-based approach showed little change---the largest cluster only consisted of 7 records. While our proposed approach still clustered 40 records of typographic digit-1s, which is much better than the other. The clustering results can be found in the supplemental material files.


\subsection{Limitations and Future Work}
We discuss two limitations of HetVis and summarize our future work.

\textbf{Scalability.} The PCA-based algorithms employed in HetVis can hardly support the analysis of data with thousands of dimensions. We need to integrate high-performance dimensionality reduction approaches and dimension recommendation approaches. As mentioned by E3, the design of the label exploration view also has difficulty in adapting to a large number of labels. We plan to optimize this view by providing an overview and recommending significant labels in the future. 

\textbf{Extensibility.} HetVis supports vector data, such as image data and tabular data. But text data and other data modalities are not supported by the current system due to different requirements of  federated learning settings. Extending to new data modalities and machine learning tasks is an interesting future work.

\section{Conclusion}
In this paper, we propose \techname, a visual analysis approach to assist identification and examination of data heterogeneity under the privacy limit of HFL. Instead of directly comparing local data and global data, we compare the output of the HFL model with a stand-alone training model. 
A contrastive clustering analysis approach is leveraged to extract heterogeneity issues from the inconsistent records identified from the output comparison. In the future, we would like to extend our system to support online tuning of HFL model. The code of our system is available at the following link: \url{https://github.com/EmmaammE/HetVis}.



\acknowledgments{
This work was supported by the National Natural Science Foundation of China (No. 62132017, 61972389). The work of Tobias Schreck has been supported by the Austrian FFG-COMET-K1 Center Pro\textsuperscript{2}Future (Products and Production Systems of the Future), Contract No.\ 881844.
}


\bibliographystyle{abbrv-doi}

\bibliography{template}
\end{document}